\pdfoutput=1
\documentclass[%
 reprint,
 amsmath,amssymb,
 prl,aps,
]{revtex4-1}

\usepackage{graphicx}
\usepackage{dcolumn}
\usepackage{bm}

\begin{document}


\title{Suppression of coherence collapse in semiconductor Fano lasers}

\author{Thorsten S. Rasmussen}
 \email{thsv@fotonik.dtu.dk}
\author{Yi Yu}%
 \email{yiyu@fotonik.dtu.dk}
 \author{Jesper Mork}
  \email{jesm@fotonik.dtu.dk}
\affiliation{%
 DTU Fotonik, Technical University of Denmark, DK-2800 Kongens Lyngby, Denmark
}%

\date{\today}

\begin{abstract}
We show that semiconductor Fano lasers strongly suppress dynamic instabilities induced by external optical feedback. A comparison with conventional Fabry-Perot lasers shows orders of magnitude improvement in feedback stability, and in many cases even total suppression of coherence collapse, which is of major importance for applications in integrated photonics. The laser dynamics are analysed using a generalisation of the Lang-Kobayashi model for semiconductor lasers with external feedback, and an analytical expression for the critical feedback level is derived.
\end{abstract}

\maketitle

Nonlinear dynamical systems with time-delayed feedback show rich and varied physics due to their infinite-dimensional nature \cite{DOYNEFARMER1982} with important examples in a large variety of disciplines, such as mechanics \cite{Hu2013}, physiology \cite{Mackey1977}, and neural networks \cite{Liao2000}. Semiconductor lasers with external optical feedback are one of the most studied examples of such delay systems, due to their wide range of applications and the serious issue of instabilities, chaos and coherence collapse arising from even extremely weak feedback \cite{Lang1980, Lenstra1985, Mork1990, Mork1992, Petermann1995}. The inherent sensitivity of these lasers towards external feedback, as well as the nature of the non-linear dynamics, remain open problems under study \cite{Sciamanna2015, Wishon2018,Huang2018, Munnelly2017, vanSchaijk2018}, and the instabilities are a particularly relevant issue in on-chip applications, due to the absence of integrated optical isolators. This has led to a number of novel solution proposals, e.g. isolators based on topological photonics \cite{Takata2018,Lu2014}, reduction of the alpha-factor \cite{Duan2018, Huang2018}, increased damping \cite{Huang2018}, and complicated laser geometries \cite{vanSchaijk2018}. Most studies have, as of yet, dealt with macroscopic lasers, while the feedback dynamics of emerging microlasers and nanolasers remain largely unexplored, with few exceptions \cite{Munnelly2017,Wang2019}. Here, we explore part of this new regime, by showing that a simple microscopic laser geometry in which one mirror is realised by a Fano resonance, providing a so-called Fano laser (FL)\cite{Mork2014, Yu2017}, is intrinsically exceedingly stable towards external optical feedback, in some cases entirely suppressing coherence collapse. The origin of the strongly enhanced stability is identified as a unique reduction of the relaxation oscillation (RO) frequency which suppresses a period-doubling route to chaos, and it is shown how the Fano laser outperforms lasers with conventional, non-dispersive mirrors by orders of magnitude in terms of feedback stability. The Fano laser is analysed using a generalisation of the traditional Lang-Kobayashi model \cite{Lang1980} for semiconductor lasers with external optical feedback. \par 
\begin{figure}[t!]
\begin{center}
\includegraphics[width = 0.45\textwidth]{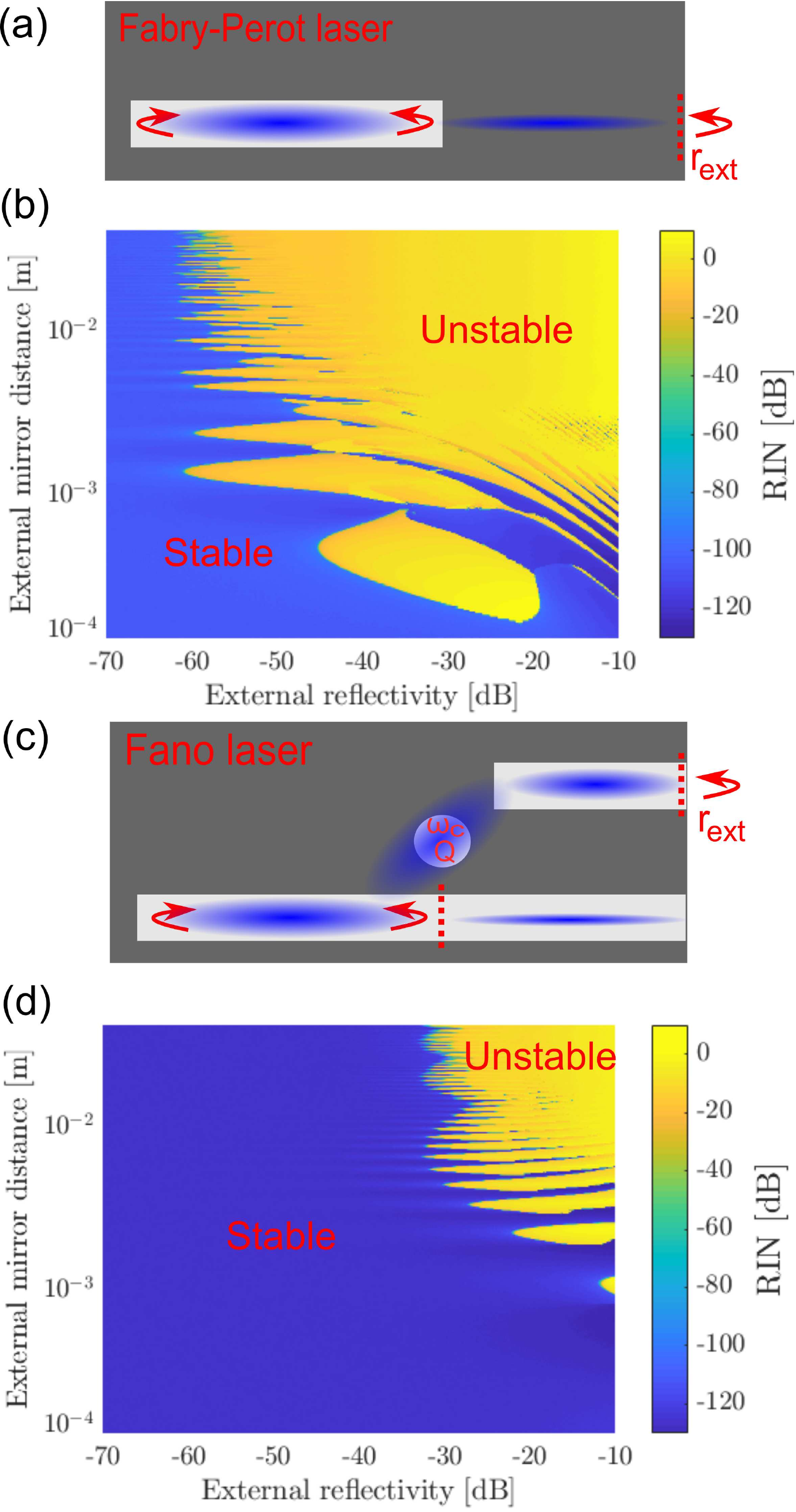}
\caption{\textbf{(a)} Schematic of Fabry-Perot laser. \textbf{(b)} Relative intensity noise (RIN) in dB as function of external reflectivity and on-chip distance to the external mirror for the FP laser.\textbf{(c)} Schematic of Fano laser. \textbf{(d)} RIN as function of external reflectivity and on-chip distance to the external mirror for the Fano laser.\vspace{-0.8cm}}
\label{fig:phasecomp}
\end{center}
\end{figure}
Figure \ref{fig:phasecomp} shows schematic representations of the Fano laser and the conventional Fabry-Perot (FP) laser, and their corresponding phase diagrams. The FL consists of a waveguide terminated at one end, which is side-coupled to a nearby cavity, from whence the Fano interference between the continuum of waveguide modes and the discrete nanocavity mode leads to a narrowband reflection peak \cite{Fan2003,Miroshnichenko2010} with bandwidth inversely proportional to the quality factor of the nanocavity. This realises a bound-mode-in-continuum \cite{Hsu2016}, which functions as a cavity mode. This configuration was experimentally realised in a photonic crystal platform \cite{Yu2017}, showing remarkable properties including pinned single-mode lasing and the first case of self-pulsing in a microscopic laser \cite{Yu2017,Rasmussen2017}, and theory suggests that its frequency modulation bandwidth is orders of magnitude larger than conventional lasers \cite{Rasmussen2018}. A comprehensive review of Fano lasers is given in \cite{Mork2019}. The photonic crystal platform itself has provided numerous promising  microscopic lasers for the photonic integrated circuits of the future \cite{Matsuo2013, Crosnier2017,Ota2017}. \par
Figures \ref{fig:phasecomp}\textbf{(b)} and \textbf{(d)} show the calculated relative intensity noise (RIN) as a function of the external reflectivity, $|r_3|^2$, and the distance to the external mirror, $L_D$, for a Fabry-Perot laser (top) and a Fano laser (bottom). The Fabry-Perot laser is described using the conventional Lang-Kobayashi model \cite{Lang1980}, while the Fano laser is modelled using a generalised version of the Lang-Kobayashi model, to be presented later. The RIN is defined as $\text{RIN} = \delta P(t) ^2/\langle P(t)\rangle ^2$, where $\delta P(t) ^2$ is the variance and $\langle P(t)\rangle$ the mean of the time-domain output power. This provides a convenient quantitative measure of the stability, with low RIN (blue) indicating stable continuous-wave (CW) states, and high RIN (yellow) corresponding to self-sustained oscillations, chaotic dynamics and coherence collapse. Despite the intricacies of these phase diagrams, the main point is clear: The Fano laser provides an extraordinary improvement in feedback stability, as shown simply by comparing the sizes of the blue and yellow regions in figure \ref{fig:phasecomp}. The critical feedback level, at which the laser is stable irrespective of the distance to the external reflector, is seen to be three orders of magnitude larger for the Fano laser compared to conventional lasers. Additionally, the Fano laser is essentially immune to feedback when the length scale reaches on-chip dimensions ($L_D \simeq 1$ mm), whereas this does not happen until $L_D \lesssim 100$ $\mu$m for the Fabry-Perot laser. Furthermore, instabilities, chaos, and coherence collapse are completely suppressed for any feedback level for the Fano laser for certain ranges of delay lengths, a characteristic not observed for Fabry-Perot lasers. The supplemental material presents a bifurcation analysis using Refs. \citenum{Tkach1986,Heil2001,Mork1988}, showing a period-doubling route to chaos for the Fano laser.  \par
The phase diagram for the Fabry-Perot laser is generated using the conventional Lang-Kobayashi model for semiconductor lasers with external feedback \cite{Lang1980}. This model has proven to work well for both Fabry-Perot and distributed feedback lasers, as well as VCSELs \cite{Petermann1995}, but in order to describe the Fano laser a generalisation is necessary. The generalisation consists of coupling the Lang-Kobayashi model to a dynamical equation for the field stored in the nanocavity\cite{Mork2014, Rasmussen2017}, in order to temporally resolve the Fano interference. This approach is also of interest for studying other coupled systems with complicated external feedback arrangements due to the generic nature of the formulation. As the output power is mainly coupled out through the cross-port \cite{Mork2014}, the feedback is assumed to originate in this port, as illustrated in figure \ref{fig:phasecomp}\textbf{(c)}. This leads to the following model equations:
\begin{align} \label{eq:Aplus}
\frac{\text{d}A^+(t)}{\text{d}t} =& \frac{1}{2}(1-i\alpha)\left(\Gamma v_g g(N) - \frac{1}{\tau_p}\right) A^+(t) \\
&+\frac{1}{\tau_{in}} A(t,\tau_D)+F_L(t) \notag \\
\frac{\text{d}N(t)}{\text{d}t} =& R_p -\frac{N}{\tau_s} -  v_g g(N) N_p
\end{align}
Here $A^+(t)$ is the envelope of the complex electric field in the laser cavity, $N(t)$ is the carrier density in the active region, $\alpha$ is the linewidth enhancement factor, $\Gamma$ is the field confinement factor, $v_g$ is the group velocity, $g(N) = g_N(N-N_0)$ is the gain, with $g_N$ being the differential gain and $N_0$ the transparency carrier density, $\tau_p$ is the photon lifetime, $\tau_{in} = v_g/(2L)$ is the laser cavity roundtrip time, with $v_g$ being the group velocity and $L$ the cavity length, $F_L(t)$ is a complex Langevin noise source, $R_p$ is the pump rate, $\tau_s$ is a characteristic carrier lifetime, and $N_p = \sigma_s |A^+(t)|^2/V_p$ is the photon density, where $V_p$ is the photon volume and $\sigma_s$ is given in Ref. \citenum{Tromborg1987}. Finally, $A(t,\tau_D)$ is related to the reflected external field, and differs for the FP laser and the FL:
\vspace{-1cm} 
\begin{align}\label{eq:Afb} \notag
\intertext{Lang-Kobayashi:}
A(t,\tau_D) =& \kappa A^+(t-\tau_D)e^{i\omega \tau_D}
\intertext{Fano Lang-Kobayashi:}
A(t,\tau_D) =& \frac{\sqrt{\gamma_c}}{r_2}A_c(t,\tau_D) -A^+(t)  \label{eq:Afb_fano}\\
\frac{\text{d}A_c(t,\tau_D)}{\text{d} t} =& (-i \delta_c-\gamma_T)A_c(t)+i\sqrt{\gamma_c}A^+(t) \label{eq:Act}\\
&-\gamma_{cp}r_3 A_c(t-\tau_D) e^{i\omega \tau_D} \notag
\end{align}
Here $\kappa = (1-|r_2|^2)r_3/|r_2|$ is the conventional definition of the external feedback coefficient \cite{Mork1992}, with $r_2$ and $r_3$ being the laser cavity and external mirror amplitude reflectivities respectively, while $\tau_D$ is the delay time of the external feedback and $\omega$ is the laser oscillation frequency. $A_c(t,\tau_D)$ is the field stored in the nanocavity, $\gamma_c$ is the field coupling rate from the waveguide to the nanocavity, $\delta_c = \omega_c - \omega$ is the detuning of the resonance frequency of the nanocavity ($\omega_c$) from the laser frequency, $\gamma_T$ is the total decay rate of the nanocavity field, and $\gamma_{cp}$ is the field coupling rate from the nanocavity into the cross-port. For the FP laser, the external feedback enters directly into equation \eqref{eq:Aplus} \cite{Tromborg1987}, while for the FL it couples to equation \eqref{eq:Aplus} through the nanocavity field, $A_c(t, \tau_D)$. This field is the solution to the conventional coupled-mode theory equation \cite{Fan2003}, but extended to self-consistently include the external feedback in equation \eqref{eq:Act}. In the calculations, parameters appropriate for microscopic lasers are used, which by itself leads to a notably lower critical feedback level than is usually observed for macroscopic Fabry-Perot lasers. Due to the cavity length being in the few-$\mu$m range, the critical feedback level is $\approx -60$ dB, which is orders of magnitude lower than the $\approx -40$ dB observed for macroscopic lasers \cite{Petermann1995, Mork1992}.  The parameters are given in the supplementary, together with an efficient iterative formulation of the model equations. \par
\begin{figure}[t!]
\begin{center}
\includegraphics[width = 0.48\textwidth]{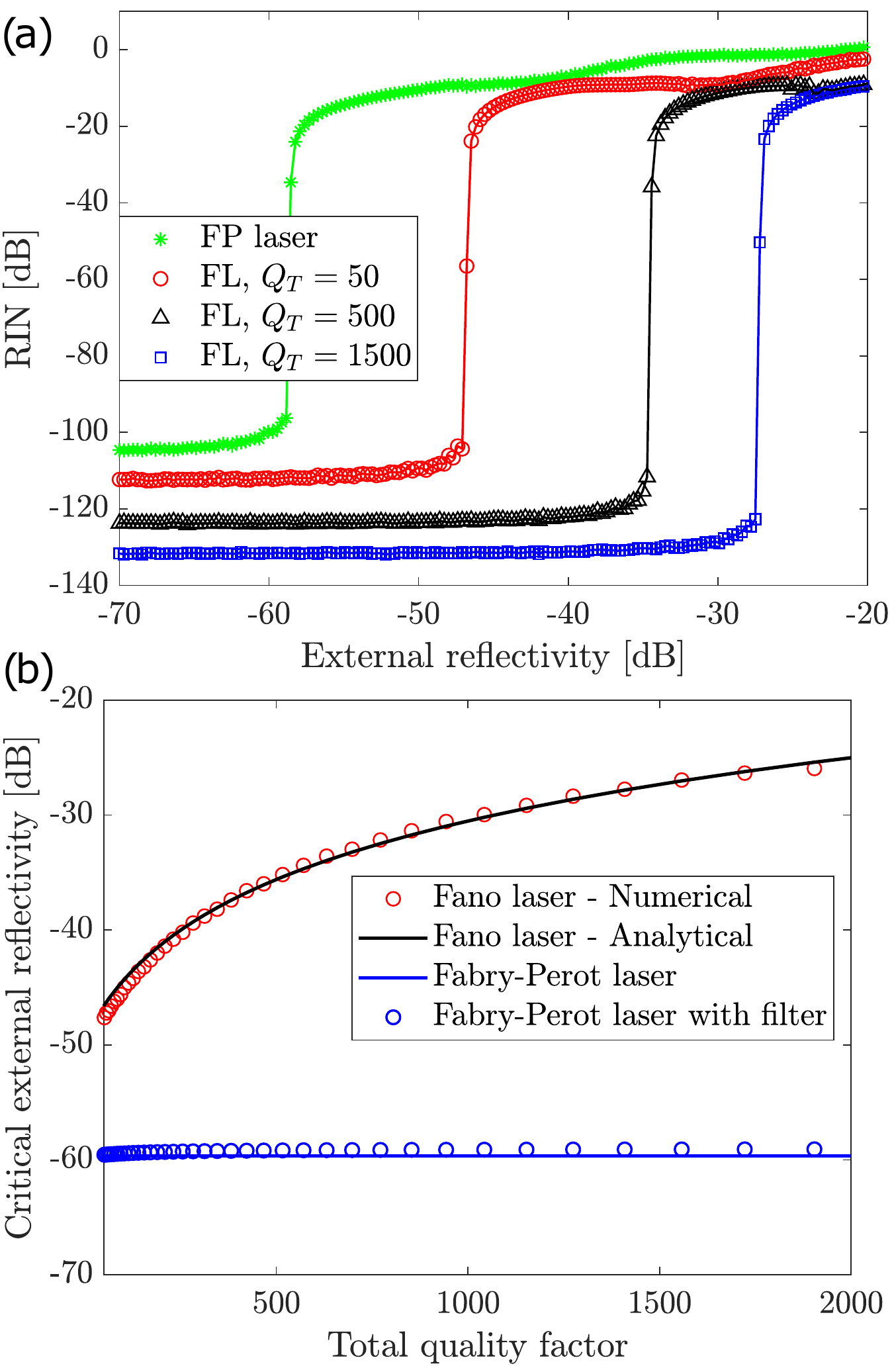}
\caption{\textbf{(a)} RIN as function of the external power reflectivity in dB with $L_D = 3 $ cm for a Fabry-Perot laser (stars) and Fano lasers with increasing nanocavity Q-factors (circles, triangles, squares). Solid lines are guides to the eye. \textbf{(b)} Numerical (red circles) and analytical (black line) critical feedback level as function of the total quality factor of the nanocavity.\vspace{-0.5cm}}
\label{fig:RIN_curve}
\end{center}
\end{figure}

We next turn to the physics responsible for the improved feedback stability of the Fano laser. The effect of the Fano mirror bandwidth is investigated in figure \ref{fig:RIN_curve}\textbf{(a)}, showing the variation of the RIN with external reflectivity, $|r_3|^2$, for a Fabry-Perot laser (circles) and Fano lasers with increasing values of the nanocavity Q-factor for $\tau_D = 1$ ns. 
The four curves show the same qualitative shape, reflecting the phase diagrams of figure \ref{fig:phasecomp}. The crucial difference between the curves, however, is the critical external reflectivity at which the onset of instability occurs, which varies by orders of magnitude. For low Q, the Fano laser RIN curve is close to the Fabry-Perot laser curve, and the critical feedback level then increases dramatically with the increase in quality factor, as shown in figure \ref{fig:RIN_curve}\textbf{(b)}. From a stability analysis one finds that in the short-cavity limit ($2Q_T/\omega \gg \tau_{in}$) the Fano laser critical external reflectivity is given by
\begin{equation} \label{eq:r3c}
|r_{3C,FL}|^2 = \left(\frac{2 Q_T\gamma_{FL}|r_2|}{\omega \sqrt{1+\alpha^2}(1-|r_2|^2)}\right)^2
\end{equation}
where $\gamma_{FL}$ is the Fano laser RO damping rate. This expression is plotted as the black line in figure \ref{fig:RIN_curve}\textbf{(b)}, showing excellent agreement with the numerical results (red circles). Equation \eqref{eq:r3c} is identical to that of Fabry-Perot lasers \cite{Tromborg1990}, except that half the laser roundtrip time ($\tau_{in}/2$) is replaced by $2Q_T/\omega$, i.e. the storage time of the energy in the nanocavity ($\tau_{p,nc}$). Thus, the improvement in feedback stability is simply
\begin{equation} \label{eq:deltar3c}
\Delta r_{3,C} \  \mathrm{[dB]} = 20\log_{10}\left(\frac{2\tau_{p,nc}}{\tau_{in}}\right)
\end{equation}
where $\Delta r_{3,C}$ is the difference in dB between the critical external power reflectivity of a Fano laser and a Fabry-Perot laser with the same parameters. This shows how the improvement in stability is an intrinsic property of the Fano mirror, independent of all parameters other than the Q-factor and the roundtrip time.\par
The appearance of the Q-factor as the governing parameter for the stability means the improvement is inversely proportional to the mirror linewidth, and as such, one might intuitively think that the Fano laser simply acts a filter for the external feedback. Such filtered feedback systems are well-studied, showing altered mode selection properties \cite{Yousefi1999}, changes to the laser dynamics \cite{Fischer2004}, a sensitive dependence on the filter width \cite{Erzgraber2007}, and phase-dependent stability improvements \cite{Tronciu2006}. The blue circles in figure \ref{fig:RIN_curve}\textbf{(b)} are a numerical calculation of the critical feedback level for a Fabry-Perot laser with filtered feedback, as function of the filter width, showing how in the range of relevant Q-factors, the filtering is essentially broadband and negligible, so that the stability improvement of the Fano laser cannot be explained simply as a filtering effect. This is the case for the entire range of delay lengths covered in figure \ref{fig:phasecomp}.\par
The improved stability can instead be understood by considering the paths to instability for lasers with feedback: Competition between allowed external cavity modes and relaxation oscillations becoming un-damped through a Hopf-bifurcation. The external cavity mode selection properties of Fano lasers in the presence of feedback differ fundamentally from conventional semiconductor lasers, and instead function as an amplified version of the improved mode selection of lasers with filtered feedback \cite{Yousefi1999}. For Fano lasers, the number of external cavity modes is strongly reduced due to the narrow bandwidth of the Fano mirror, which significantly increases the threshold gain of modes that are separated in frequency by more than $1/\gamma_T$. In this case an effective C-parameter \cite{Petermann1995} for the Fano laser may be estimated as
\begin{equation}
C_{FL} = \kappa_{FL}  \frac{\tau_D}{ \frac{2Q_T}{\omega}+\tau_{in}}\sqrt{1+\alpha^2}
\end{equation}
where $\kappa_{FL}=r_3(1-|r_2|) = r_3\gamma_{cp}/\gamma_T$ acts as an effective feedback parameter for the Fano laser. In comparison to lasers with filtered feedback, the filtering of external cavity modes is much stronger, because a detuning in frequency changes the internal reflectivity of the laser, rather than simply the feedback strength, leading to a much larger threshold gain separation between neighbouring external cavity modes. This extreme sensitivity to frequency changes ensures that a single, dominant external cavity mode is selected, so that the critical feedback level is uniquely determined by the location of the first Hopf bifurcation, and the location of this Hopf bifurcation is what is determined by Eq. \eqref{eq:r3c}. \par
The origin of the scaling with the Q-factor in the expression for the critical feedback level is a unique reduction in the relaxation oscillation frequency for the Fano laser compared to Fabry-Perot lasers without notable change of the RO damping rate. The benefit of this reduction in terms of feedback stability is that the instability is born from ROs becoming un-damped. Thus, if the frequency to damping rate ratio is reduced, fewer oscillations take place before a perturbation decays back to the steady-state. Because of this, a larger level of feedback is necessary to drive the instability, leading to the feedback stability scaling with the Q-factor. This type of reduction of the RO frequency is particularly efficient for suppressing coherence collapse, because the route to chaos is of the period-doubling form. As the temporal period is controlled by the RO frequency, a lower frequency means a longer period, requiring stronger feedback to sustain the oscillations through the longer and longer periods in order to drive the laser towards chaos. We believe that due to the generic nature of the FL equations and the stability mechanism, this type of behaviour could also be exploited to suppress chaotic dynamics in other delay systems if the natural frequency of the system can be engineered, in particular for systems governed by period-doubling routes to chaos as is the case here.\par
\begin{figure}[t!]
\begin{center}
\includegraphics[width=0.5\textwidth]{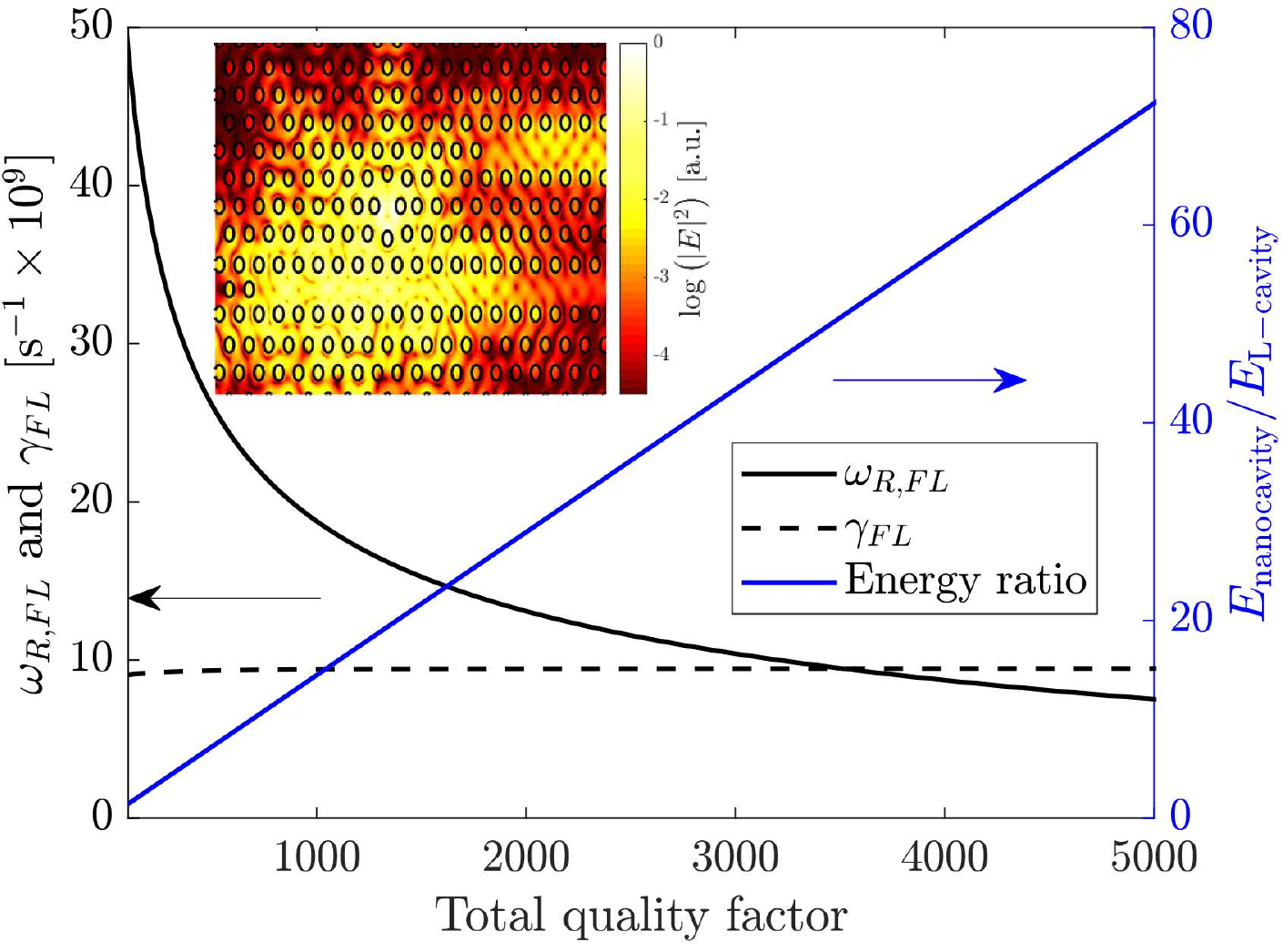}
\caption{Fano laser relaxation oscillation angular frequency (full black, left axis) and damping rate (dashed black, left axis), and ratio of energy stored in the nanocavity relative to the laser cavity (blue, right axis), all as function of the total quality factor. Inset: Simulated field distribution during lasing (logarithmic scale). Note the bright spots in the nanocavity.}
\label{fig:ratio_plot}
\end{center}
\end{figure}

The reduction of the RO frequency with Q-factor arises because of a longer effective photon lifetime of the system, as the high-Q nanocavity stores a significant amount of the intensity during lasing (see inset in figure \ref{fig:ratio_plot}). As the Q-factor increases, this amount increases, as shown in figure \ref{fig:ratio_plot} (right axis, blue curve), which in turn means that the interaction between photons and free carriers in the laser cavity is weaker, leading to a smaller relaxation oscillation frequency (full black, left axis), similar to reducing the photon number or increasing the photon lifetime of a conventional laser \cite{CC}. \par
A small-signal analysis of the feedback-free FL equations yields the FL RO frequency as
\begin{align}
\omega_{R,FL}^2 &= \omega_R^2\left(1-\frac{1}{1+\frac{\omega \tau_{in}}{2Q_T}}\right) \label{eq:wr_approx}
\end{align}
Here $\omega_R$ is the corresponding RO frequency for a Fabry-Perot laser with the same parameters \cite{CC}. The connection to equation \eqref{eq:r3c} and \eqref{eq:deltar3c} is clear, since in the short-cavity limit we have 
\begin{equation}
\frac{\tau_{p,nc}}{\tau_{in}} =\frac{\omega_R^2}{\omega_{R,FL}^2}
 \end{equation}
showing how the reduction of the RO frequency explains the improved feedback stability. \par
Here, one should appreciate the fundamental difference between Fano and FP lasers. As additional energy is stored in the nanocavity due to increasing Q-factor, the RO frequency decreases until eventually the FL approaches an over-damped regime ($Q_T \gtrapprox 3500$).
The absence of relaxation oscillations can be interpreted as a transition of the FL from a class B laser towards a class A laser \cite{Tredicce1985}, i.e. a system of lower dimensionality, which leads to an intrinsic suppression of chaos as the quality factor increases.  For FP lasers, in contrast, the critical feedback level is determined exclusively by the damping rate \cite{Petermann1995}, as exemplified by stability improvements due to a short carrier lifetime due to growth defects in Ref. \citenum{Huang2018}. Figure \ref{fig:ratio_plot} shows that despite the strong variation of the RO frequency, the Fano laser damping rate only changes marginally with Q. In contrast, the damping rate in conventional lasers increases approximately with the square of the RO frequency, showing how the mechanism of improved feedback stability in Fano lasers is fundamentally different. \par 
In conclusion, it has been shown how semiconductor Fano lasers intrinsically suppress dynamic instabilities induced by exposure to external optical feedback. A generalisation of the Lang-Kobayashi model was employed to study the laser dynamics. Using this, the feedback stability was analytically and numerically shown to scale with the quality factor of the nanocavity, due to a unique reduction of the relaxation oscillation frequency, as well as large gain separation of external cavity modes due to the highly dispersive Fano mirror. For realistic designs, the Fano laser outperforms conventional Fabry-Perot lasers by orders of magnitude in terms of the critical external feedback level. In many cases, coherence collapse is even entirely suppressed, in contrast to the conventional classification of semiconductor lasers with feedback, demonstrating fundamentally new and different underlying dynamics arising from the Fano mirror. Due to the general nature of the problem, this stability mechanism may be exploitable for chaos suppression in similar delay systems.

\begin{acknowledgments}
Authors gratefully acknowledge funding by Villum Fonden via the NATEC Center of Excellence (Grant 8692). This project has received funding from the European Research Council (ERC) under the European Union's Horizon 2020 research and innovation programme (grant agreement number 834410 — FANO).
\end{acknowledgments}

%

\end{document}



\title{Suppression of coherence collapse in semiconductor Fano lasers - Supplementary}

\author{Thorsten S. Rasmussen}
 \email{thsv@fotonik.dtu.dk}
\author{Yi Yu}%
 \email{yiyu@fotonik.dtu.dk}
 \author{Jesper Mork}
  \email{jesm@fotonik.dtu.dk}
\affiliation{%
 DTU Fotonik, Technical University of Denmark, DK-2800 Kongens Lyngby, Denmark
}%

\date{\today}
             
\maketitle             
This supplementary outlines the computational method used to generate the numerical results presented in the main paper, lists the parameters used, shows a brief bifurcation analysis of the route to chaos, and demonstrates the Fano mirror dependence of the Q-factor, and how this affects the relaxation oscillation damping rate. 
\section*{Computational model}
As mentioned in the main text, the phase diagrams for the Fabry-Perot laser are generated using the conventional Lang-Kobayashi (LK) model for semiconductor lasers with external feedback \cite{Lang1980}, while the Fano laser diagrams are computed using a generalised version of the LK model.  The LK model has proven to work well for both Fabry-Perot and distributed feedback lasers, as well as VCSELs \cite{Petermann1995}, but in order to describe the Fano laser a generalisation is necessary. The generalisation consists of coupling the conventional LK model to a dynamical equation for the field stored in the nanocavity, in order to temporally resolve the Fano interference, as in the previously developed FL model \cite{Mork2014, Yu2017, Rasmussen2017}. As shown in \cite{Mork2014} the ratio of power between the upper (denoted cross-port) and lower (denoted through-port) output port is generally large when operating near resonance ($\gtrsim 10$), meaning that the useful output will be in this channel. Based on this, we assume that the external feedback originates in this port, which, when combined with assuming weak external feedback ($|r_3| \ll 1$), leads to the model equations $(1)-(5)$ given in the main text. As a validation of this modelling approach, it is straight-forward to show that in the limit where the Fano mirror bandwidth is very large and the nanocavity field can be adiabatically eliminated, this formulation correctly reduces to the conventional form of the Lang-Kobayashi model. In order to reduce computational demands, we have developed a highly efficient iterative formulation of this model, which is presented here in the supplementary.\\
The version of the LK-model in the main text is derived using a first-order expansion of the steady-state oscillation condition as explained in \cite{Tromborg1987}. In short, one writes up a field equation at a reference plane next to the right mirror of the laser cavity, e.g. 
\begin{equation} \label{eq:fieldeq}
E^+(\omega,z_0)= r_{LS}E^-(\omega,z_0) + F_\mathrm{noise}(\omega)
\end{equation}
where $r_{LS} = r_L(\omega_s,N_s)$ is the effective left-mirror reflectivity, which includes the roundtrip gain and phase of the entire cavity and $E^\pm(z_0)$ represent the right(+) and left(-) propagating field amplitudes at the reference plane ($z = z_0$), while $F_\mathrm{noise}(\omega)$ accounts for spontaneous emission throughout the cavity. The steady-state solutions ($\omega_s,N_s$) are found by solving the oscillation condition
\begin{equation}
r_L(\omega_s,N_s) r_R(\omega_s) = 1
\end{equation}
where $r_R(\omega_s)$ is an effective right mirror reflectivity also including reflections from the external mirror. The effective left reflectivity $r_{LS}$ is given by 
\begin{equation} \label{eq:rleq}
r_{LS} = r_1 \exp[-2iLk(\omega_s,N_s)]
 \end{equation} 
where $r_1$ is the left mirror amplitude reflectivity, $L$ is the cavity length and $k(\omega_s,N_s)$ is the complex wavenumber accounting for roundtrip gain/loss and phase. To obtain the differential equation form of Eq. (1) in the main text, equation \eqref{eq:fieldeq} is divided by $r_L(\omega_s,N_s)$ and the term $1/r_L(\omega_s,N_s)$ is expanded to first-order in frequency and carrier density. The Fourier transform of this then yields Eq. (1) in the main text by neglecting the frequency dependence of the left mirror and the gain. If one instead limits the first-order expansion to the wavenumber, i.e. $k(\omega_s,N_s)$ in \eqref{eq:rleq}, one instead obtains a convenient difference equation after taking the Fourier transform, which can be readily implemented numerically. This has the additional benefit that it actually provides \textbf{better} accuracy than the differential equation model, since only the wavenumber is expanded, and not the entire exponential of eq. \eqref{eq:rleq}. Using this approach, the equation for $A^+(t)$ takes the following form:
\begin{align}
A^+(t+\tau_{in}) = &r_{LS}A^-(t,\tau_D)\exp\left[\frac{\tau_{in}}{2} G(N-N_s)(1-i\alpha)\right] \notag \\
&+ \tau_{in} F_L(t)
\end{align}
Here $A^-(t,\tau_D)$ represents the left-propagating field at the reference plane, $\tau_{in}$ is the L-cavity roundtrip time, and $G = \Gamma v_g g_N$, so that the exponential represents the deviation from the expansion point of the roundtrip gain and phase. For the conventional laser, we have simply that
\begin{equation}
r_{LS} A^-(t,\tau_D) = A^+(t) +\kappa A^+(t-\tau_D)e^{i\omega \tau_D}
\end{equation} 
so that the evolution equation becomes
\begin{align} \label{eq:AP_FP}
A^+(t+\tau_{in}) = &\exp\left[\frac{\tau_{in}}{2} G(N-N_s)(1-i\alpha)\right]\times \big\{A^+(t) \notag \\
& +\kappa A^+(t-\tau_D)e^{i\omega \tau_D}\big\} + \tau_{in} F_L(t)
\end{align}
as given in \cite{Mork1988}. This practical form allows for iterative stepping through the temporal evolution of the field envelope with resolution given by the round-trip time. Since the laser cavity lengths studied here are so short (few microns), this yields a temporal resolution of $\approx 100$ fs, which is sufficient to study even very short external cavities ($\tau_D \gtrsim 1$ ps, on-chip ). The Langevin noise source is generated as described in \cite{Mork1988}, with the interpretation that $\tau_{in}F_L(t)$ corresponds to the integrated contribution from a single time step $\tau_{in}$. This approximation is valid when there are many spontaneous emission events in a single time step, i.e. $R_{sp}\tau_{in} \gg 1$.  For the Fano laser, the same approach is adopted, except that the left-propagating field is now the field back-coupled from the nanocavity, so that the evolution equation takes the form
\begin{align} \label{eq:AP_FL}
A^+(t+\tau_{in}) = &r_{LS}\sqrt{\gamma_c}A_c(t,\tau_D)\exp\left[\frac{\tau_{in}}{2} G(N-N_s)(1-i\alpha)\right] \notag \\
&+ \tau_{in} F_L(t)
\end{align}
where $A_c(t,\tau_D)$ is the solution to equation Eq. (5) in the main text. The differential equation for the carrier density can also be put on a similar iterative form by utilising a first-order expansion:
\begin{equation} \label{eq:N_FP}
N(t+\tau_{in}) \simeq N(t)+\tau_{in} \left.\frac{\text{d}N(t)}{\text{d}t}\right|_t
\end{equation}
with $\frac{\text{d}N}{\text{d}t}$ given by Eq. (2) of the main text. This is a reasonable approximation, since the assumption here is that the carrier density is slowly varying on the scale of the round-trip time ($\approx 100$ fs). 
The final task is to deal with the form of Eq. (5) in the main text. Since the carrier density and L-cavity field vary with the temporal resolution $\tau_{in}$, they may be taken as constant parameters from time $t$ to $t+\tau_{in}$, which then permits an analytical solution of Eq. (5) in the main text, which can be time-stepped in a straight-forward manner:
\begin{align} \label{eq:AC_FL}
A_c(t+\tau_{in}) = &\frac{-r_3 \gamma_{cp}}{i\delta_c + \gamma_T} A_c(t-\tau_D)e^{i\omega \tau_D} +  \frac{r_2}{\sqrt{\gamma_c}} A^+(t)\notag\\
&+ \bigg[A_c(t)+\frac{r_3 \gamma_{cp}}{i\delta_c + \gamma_T} A_c(t-\tau_D)e^{i\omega \tau_D} \\
&-  \frac{r_2}{\sqrt{\gamma_c}} A^+(t)\bigg] \times \exp[-(i\delta_c + \gamma_T)\tau_{in}] \notag
\end{align}
Note here that in the limit $\gamma_T\tau_{in} \ll 1$ this is equivalent to using 
\begin{equation}
A_c(t+\tau_{in}) = A_c(t) + \frac{\text{d}A_c(t)}{\text{d}t}\tau_{in}
\end{equation}
as was utilised for the carrier density, and that for the combinations of $Q$ and $L$ studied here this is in practice always fulfilled. Thus, equation \eqref{eq:AP_FP} and \eqref{eq:N_FP} constitute the evolution equations for the FP laser. The carrier density and field amplitude corresponding to the minimum linewidth mode are used as initial conditions, in order to accurately portray the relevant physical system. These values are determined as described in \cite{Mork1992}. For the Fano laser, the evolution equations are equation \eqref{eq:AP_FL}, \eqref{eq:N_FP}, and \eqref{eq:AC_FL}. The simulation time is set to 500 ns to ensure that long-term behaviour is accurately captured, avoiding the impact of slow transients. 
The following parameter values are used throughout the calculations, unless otherwise specified: $\omega_c = 2\pi c/(1.55 \mu \mathrm{m}), \alpha = 5, r_2 = 0.94, \Gamma = 0.01, v_g = c/n_g, n_g = 3.5, g_N = 5 \times 10^{-16} \mathrm{m}^{-2}, N_0 = 5 \times 10^{21} \mathrm{m}^{-3}, \tau_s = 0.28 \mathrm{ns}, \tau_p = 0.91 \mathrm{ps}$, $Q_T = 750$, and $R_P = 1.5 R_{P,th}$, i.e. 1.5 times the threshold pump rate. The volume is $V = L A$, where $L = 4.88 \mu \mathrm{m}, A = 0.20 \mu \mathrm{m}^2$. The decay rates are related to the quality factors as $\gamma_x = \omega_c/(2Q_x)$. For a total quality factor of 750, the corresponding quality factors are $Q_c  = 780, Q_v = 10^5, Q_p = 1.5 \times 10^4$, and the decay rates are related to the quality factors as $\gamma_x = \omega_c /(2Q_x)$.
\section*{Route to chaos and dynamical trajectories for different feedback levels}
\begin{figure}[t!]
\begin{center}
\includegraphics[width = 0.45\textwidth]{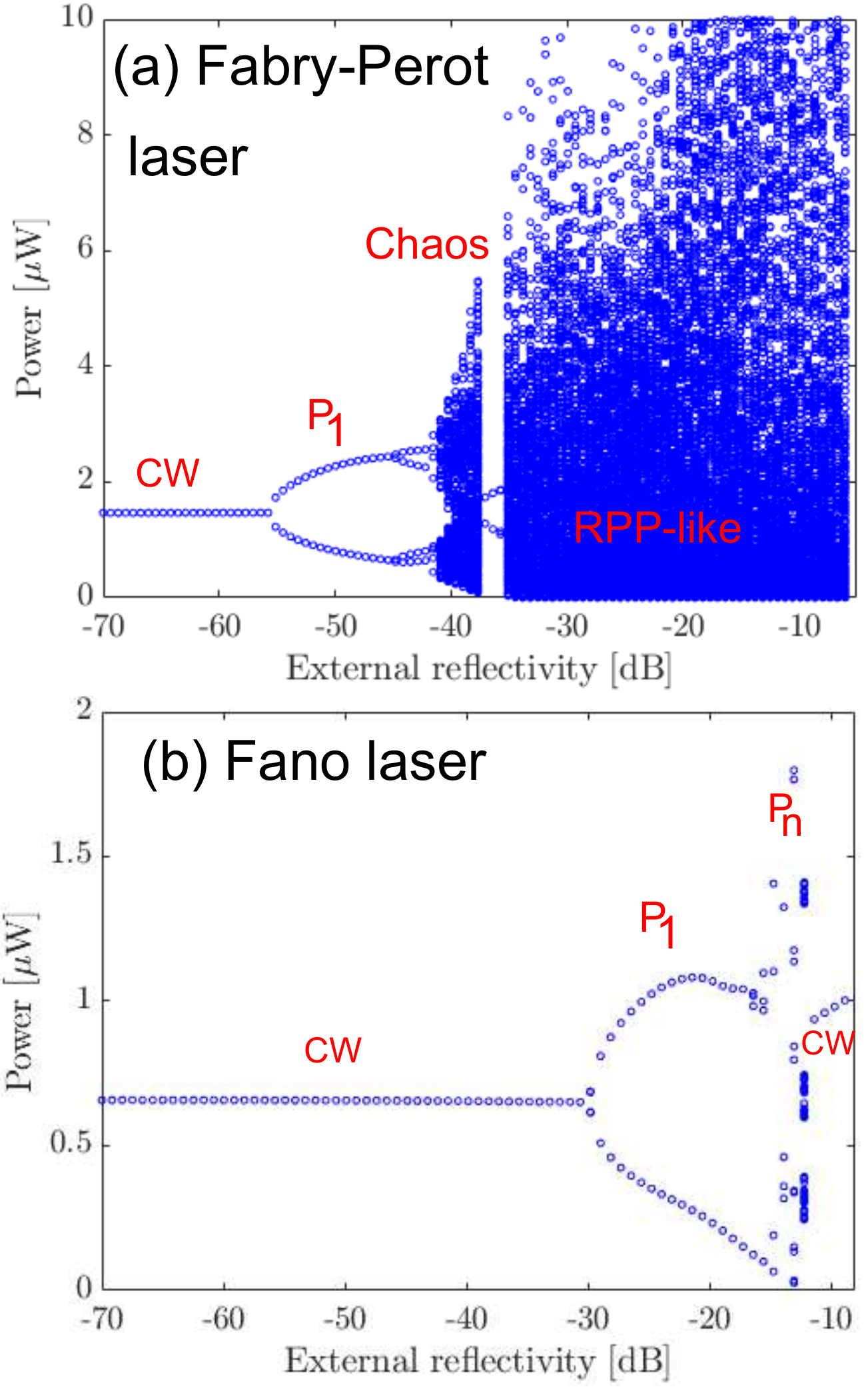}
\caption{\textbf{(a)} Bifurcation diagram for the Fabry-Perot laser with $L_D = 4.3$ mm. Markers represent local extrema in the time trace of the power output.  Red labels show the dynamical states, with $\mathrm{P_1}$ indicating a period-one oscillation, and RPP being regular pulse packages.\textbf{(b)} Bifurcation diagram for the Fano laser with $L_D = 4.3$ mm. $\mathrm{P_n}$ is an orbit of period n. See example trajectories in supplementary.}
\label{fig:bifurcations}
\end{center}
\end{figure}
The underlying dynamics of figure 1 in the main text can be investigated by sweeping along the horizontal lines, i.e. choosing a delay length and sweeping the reflectivity, which yields the bifurcation diagrams in figure \ref{fig:bifurcations}, directly demonstrating the dramatic difference between the conventional laser and the Fano laser. Here a single-valued power corresponds to a CW state, while multiple distinct values correspond to periodic orbits of order $n$, where $n$ is the number of distinct power values. The eventual total smearing out of values observed for the FP laser for increasing reflectivity corresponds to the transition to chaotic oscillations and coherence collapse.\par 
Figure \ref{fig:bifurcations} clearly illustrates the qualitative difference between FP lasers and the FL. Here, the first instability is for both lasers related to a Hopf bifurcation, where relaxation oscillations become un-damped \cite{Mork1990,Petermann1995}, leading to the characteristic branching shape in the bifurcation diagram. From the first bifurcation, however, the FL provides highly distinct dynamics. The FP laser generally transitions first into a regime of true chaotic oscillations, and then, as the feedback level is increased, enters a regime of quasiperiodic states similar to the regular pulse packages observed in Ref. \citenum{Heil2001}, for short delay lengths, as indicated by the labels in figure \ref{fig:bifurcations}. The dynamics are highly complicated due to the interplay between RO instabilities and hopping between external cavity modes, and vary in particular with the delay length \cite{Petermann1995}. For the Fano laser, however, the coherence collapse is in many cases completely suppressed. Some parameter combinations even yield large ranges of stable CW operation after the onset of the first instability without ever permitting coherence collapse, as in figure \ref{fig:bifurcations}\textbf{(b)}. Here the FL undergoes a period-doubling route towards chaos, but returns to a stable CW state before experiencing coherence collapse. Generally the period-doubling route to chaos appears to be ubiquitous for the FL, and this is crucial for efficiently suppressing coherence collapse, as explained in the main text. This behaviour is in stark contrast to the well-known classification of operation regimes in conventional lasers with feedback, where the stabilisation (regime V) follows after coherence collapse (regime IV) \cite{Tkach1986}. 
\begin{figure}[t!]
\begin{center}
\includegraphics[width=0.315\textwidth]{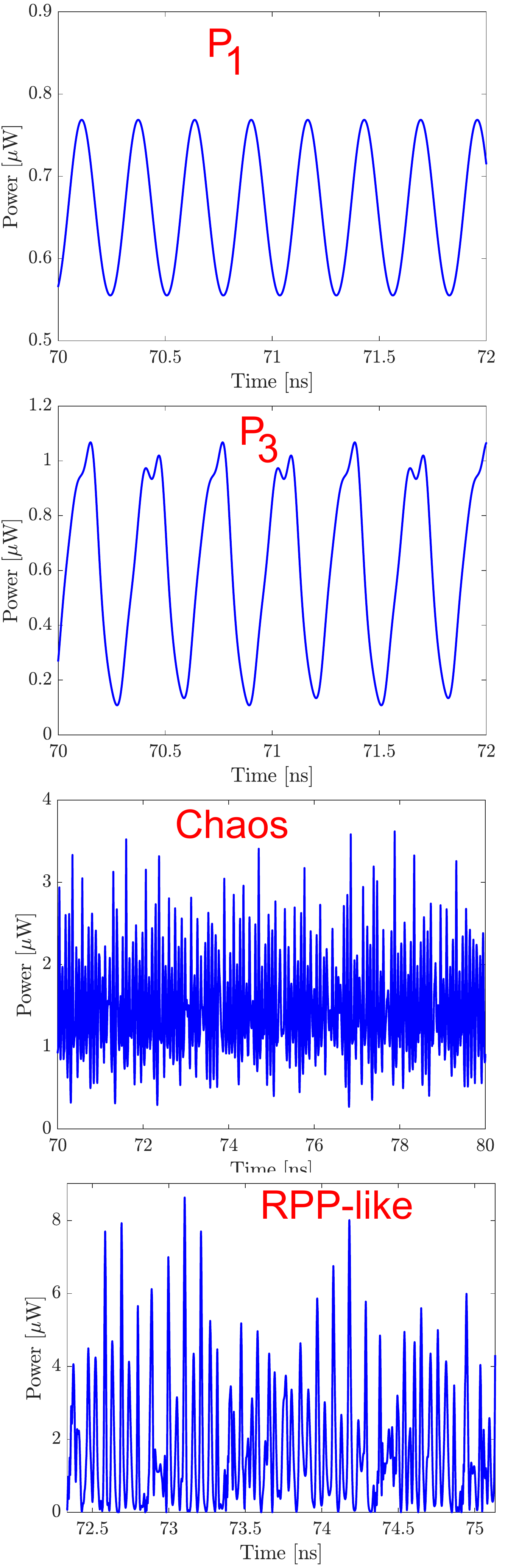}
\caption{Examples of the trajectories specified in the bifurcation diagrams in figure 2 of the main text. First from the top: Period one. Second: Period three. Third: Chaotic oscillations. Fourth: Rapid pulse packages. }
\label{fig:trajectories}
\end{center}
\end{figure}
The labels in figure \ref{fig:bifurcations} are clarified here, by showing the corresponding calculated time traces of the output power in figure \ref{fig:trajectories}. This includes a period-one ($\mathrm{P_1}$) orbit, i.e. a periodic trajectory with a single radiofrequency component, as well as a period-three ($\mathrm{P_3}$) orbit from the FL bifurcation diagram, a chaotic orbit and an example of behaviour similar to the rapid pulse packages (RPP) observed in Ref. \citenum{Heil2001}, both from the FP laser. Note the regularly spaced pulses in the last trajectory, which are separated by the round-trip time of the external cavity ($\simeq 100$ ps), and only occur for short external cavities, i.e. $\tau_D\omega_R \lesssim 1$.
For reference, the output power in the cross-port is calculated as \cite{Rasmussen2018}
\begin{equation}
P_x = 2\epsilon_0 ng c \gamma_{cp}|A_c|^2
\end{equation}
\section*{Fano mirror Q-factor dependence and damping of relaxation oscillations}
\begin{figure}[t!]
\begin{center}
\includegraphics[width=.48\textwidth]{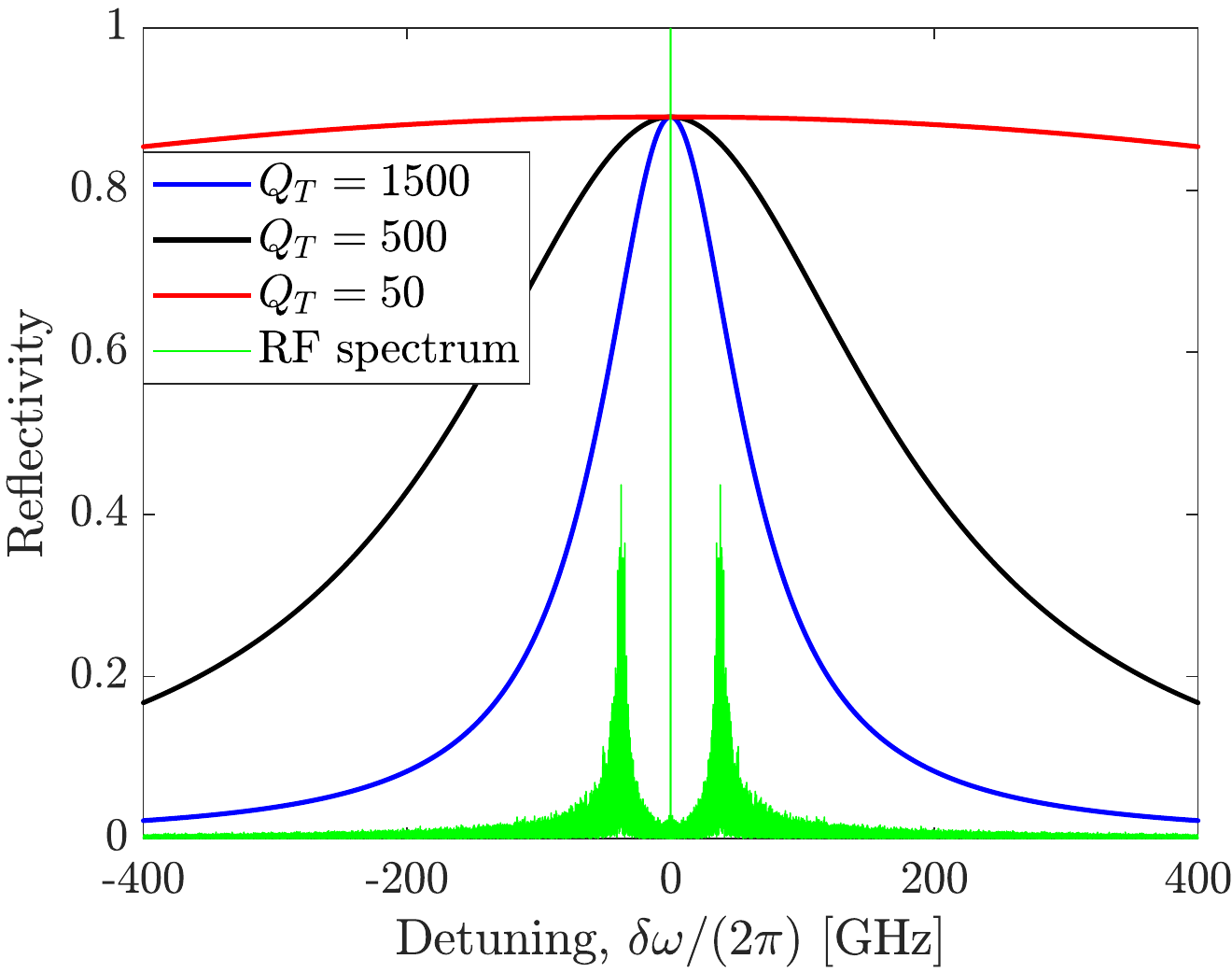}
\caption{Fano mirror power reflectivity profiles for increasing Q-factors. The green curve represents an example of a normalised lasing RF spectrum near the first Hopf bifurcation.}
\label{fig:mirror}
\end{center}
\end{figure}

The complex reflectivity of the Fano mirror can be derived using coupled-mode theory \cite{Fan2003}. In the simplest case, i.e. in the absence of a partially transmitting element, this yields a Lorentzian reflection profile \cite{Mork2014}
\begin{equation}
r_2(\omega,\omega_c) = \frac{i\gamma_c}{i(\omega_c-\omega)+\gamma_T}
\end{equation}
This shows directly how the mirror linewidth scales inversely with the Q-factor of the nanocavity, since $\gamma_T = \omega_c/(2Q_T)$. Figure \ref{fig:mirror} illustrates this, showing the power reflectivity for different values of the quality factor.  Overlaid is a simulated RF spectrum of a laser near the first Hopf bifurcation, showing how the side-peaks at the relaxation oscillation frequency are weakly damped by the Fano mirror. The damping itself scales with the relaxation oscillation frequency, so that the reduction of this frequency by the energy storage in the nanocavity in turn decreases the damping, leading to only small changes in the damping rate with Q-factor. This can be seen in figure 3 of the main text, and confirms that the additional damping of the mirror is not nearly sufficient to explain the observed increase in the critical feedback level with Q-factor, which instead is a consequence of the unique reduction of the relaxation oscillation frequency. 
%